\documentclass{sigchi}

\CopyrightYear{2020}
\setcopyright{acmlicensed}
\doi{https://doi.org/10.1145/3313831.XXXXXXX}
\isbn{978-1-4503-6708-0/20/04}
\conferenceinfo{UIST'20,}{October  20--23, 2020, Minneapolis, MN, USA}
\acmPrice{\$15.00}

\usepackage{balance}       
\usepackage{graphics}      
\usepackage[T1]{fontenc}   
\usepackage{txfonts}
\usepackage{mathptmx}
\usepackage[pdflang={en-US},pdftex]{hyperref}
\usepackage{color}
\usepackage{booktabs}
\usepackage{textcomp}

\usepackage{microtype}        
\usepackage{ccicons}          

\usepackage{todonotes}

\def\plaintitle{Dynamic Difficulty Adjustment via Fast User Adaptation}

\def\emptyauthor{}
\def\plainkeywords{Dynamic difficulty adjustment; deep learning; meta-learning.}

\makeatletter
\def\url@leostyle{%
  \@ifundefined{selectfont}{
    \def\UrlFont{\sf}
  }{
    \def\UrlFont{\small\bf\ttfamily}
  }}
\makeatother
\def\pprw{8.5in}
\def\pprh{11in}

\definecolor{linkColor}{RGB}{6,125,233}
\hypersetup{%
  pdftitle={\plaintitle},
  pdfauthor={\emptyauthor},
  pdfkeywords={\plainkeywords},
  pdfdisplaydoctitle=true, 
  bookmarksnumbered,
  pdfstartview={FitH},
  colorlinks,
  citecolor=black,
  filecolor=black,
  linkcolor=black,
  urlcolor=linkColor,
  breaklinks=true,
  hypertexnames=false
}

\begin{document}

\title{\plaintitle}

\numberofauthors{2}
\author{%
  \alignauthor{Hee-Seung Moon\\
    \affaddr{Yonsei University}\\
    \affaddr{Incheon, Korea}\\
    \email{hs.moon@yonsei.ac.kr}}\\
  \alignauthor{Jiwon Seo\\
    \affaddr{Yonsei University}\\
    \affaddr{Incheon, Korea}\\
    \email{jiwon.seo@yonsei.ac.kr}}\\
}

\maketitle

\begin{abstract}
  Dynamic difficulty adjustment (DDA) is a technology that adapts a game's challenge to match the player's skill. It is a key element in game development that provides continuous motivation and immersion to the player. However, conventional DDA methods require tuning in-game parameters to generate the levels for various players. Recent DDA approaches based on deep learning can shorten the time-consuming tuning process, but require sufficient user demo data for adaptation. In this paper, we present a fast user adaptation method that can adjust the difficulty of the game for various players using only a small amount of demo data by applying a meta-learning algorithm. In the video game environment user test (n=9), our proposed DDA method outperformed a typical deep learning-based baseline method.
\end{abstract}

\ccsdesc[500]{Human-centered computing~User models}
\ccsdesc[300]{Computing methodologies~Neural networks}
\ccsdesc[300]{Applied computing~Computer games}

\keywords{\plainkeywords}

\printccsdesc

\section{Introduction}
Difficulty balancing is a key element in game development because players easily become bored or frustrated when the games are too easy or difficult for them. Dynamic difficulty adjustment (DDA) is a method for adapting the difficulty of a game according to the player's ability to provide continuous motivation to the player. Various studies in the HCI field~\cite{baldwin2014effect, constant2019dynamic, denisova2019player} have revealed that DDA has positive effects, such as increasing the immersion~\cite{denisova2015adaptation} and long-term motivation of players~\cite{pfau2020enemy}.

Several studies have been conducted on how to implement DDA~\cite{he2010dynamic, yin2015data, melhart2019your, wu2019strength}. One of the most straightforward but powerful methods is to increase or decrease the game's strength index, e.g., the in-game parameters or the AI level, according to the player's in-game performance. However, the parameter adjustment method requires careful tuning, which is time-consuming. With the development of deep learning technology in various fields~\cite{yu2018one, moon2019observation, moon2019prediction, wu2020predicting}, it is expected that the shortcomings of conventional DDA methods can be overcome by using deep neural networks. In~\cite{pfau2020enemy}, a method was proposed for adapting the game challenge to the player by generating an enemy agent based on a player model trained using the player's actual movement and strategy. This method outperformed conventional DDA in several subjective metrics but required adequate data acquisition process because the player model had to be newly trained for each player.

In this paper, we propose a novel DDA approach referred to as fast user adaptation based on deep neural networks that can quickly adapt to a player's capabilities with a small amount of play data. In order to use the sparse user demo data effectively, we employ the \textit{model-agnostic meta-learning (MAML)} algorithm~\cite{finn2017model}. Meta-learning is a method that focuses on fast adaptation to various tasks, i.e., the generalization of network parameters, to make it easy to respond to new unseen tasks. We apply this meta-learning concept to create a DDA model that quickly adapts to new players.

\begin{figure}[b!]
\centering
  \includegraphics[width=0.92\columnwidth]{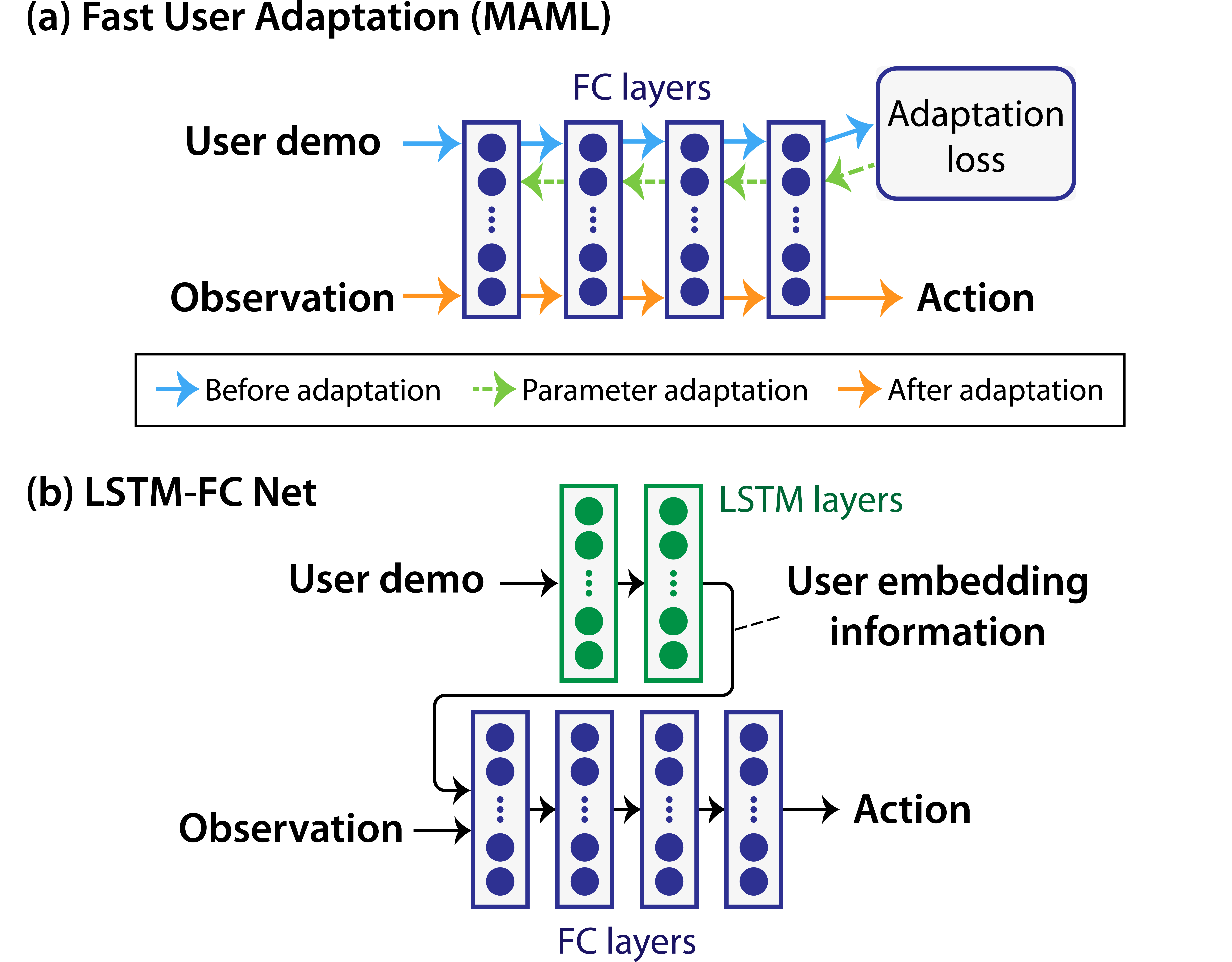}
  \caption{Overview of the DDA network models we implemented. (a) Fast user adaptation. (b) LSTM-FC Net.}~\label{fig:1}
\end{figure}

\begin{figure*}[t!]
\centering
  \includegraphics[width=2\columnwidth]{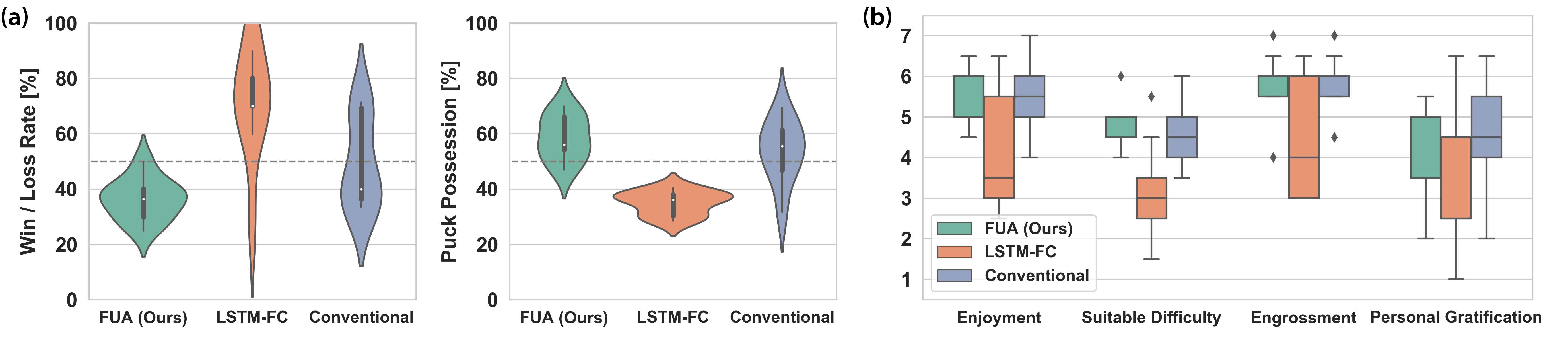}
  \caption{(a) Objective evaluation results and (b) subjective evaluation results of each DDA method.}~\label{fig:2}
\end{figure*}

\section{Fast User Adaptation}
The fast user adaptation method we present is to modify the MAML algorithm~\cite{finn2017model} to train a model that can quickly respond to different users (Figure~\ref{fig:1}(a)). When dividing the training data obtained from various tasks into \(D_{demo}\) and \(D_{valid}\), the MAML method first updates the network parameter \(\theta\) in a few gradient steps calculated using \(D_{demo}\), and trains to minimize the loss of \(D_{valid}\) calculated with the updated \(\theta\). This training method can be expressed in the following equation

\[ \min_{\theta} \sum_{T} L(\theta - \alpha \nabla_\theta L(\theta, D_{demo}^T), D_{valid}^T), \]

where \(L(\theta, D^T)\) denotes the loss value when data \(D^T\), obtained from task \(T\), is fed into the model with parameter \(\theta\). Our fast user adaptation method applies the MAML algorithm, where in place of using training data from various tasks (\(D^T\)), we use data from various players (\(D^P\)) instead. Similar to~\cite{pfau2020enemy}, we hypothesize that a DDA that makes players encounter agents whose behavior and strategy are similar to themselves can boost player motivation effectively. Therefore, our DDA method is intended to make an agent quickly learn the player's movements so that the player faces an agent who plays similarly to himself/herself.

\section{Experiment Details}
For the user test, we developed a virtual \textit{Air Hockey} game environment where two players compete with their respective strikers and a single puck on a slippery surface. A player can freely move the striker within his/her area and score points by hitting the puck and putting it inside the opponent's goal. We conducted a user test that confronts participants with DDA-applied agents in this \textit{Air Hockey} game environment.

To validate our fast user adaptation model, we implemented two baseline DDA methods: another data-driven approach utilizing neural networks, and a conventional DDA approach. For the data-driven baseline, referred to as \textit{LSTM-FC Net}, we implemented a neural network model incorporating long short-term memory (LSTM) layers that can extract the user embedding information, e.g., users' proficiency, from user demo data, and fully connected (FC) layers that output appropriate actions based on the current game state and the embedding information (Figure~\ref{fig:1}(b)). For the conventional DDA baseline, we generated agents corresponding to progressive levels of difficulty from 1 to 9. The level of difficulty was increased or decreased depending on the player's win or loss.

In detail, our DDA network consists of four FC layers with 80 hidden units, and the LSTM-FC Net consists of two LSTM layers with 10 hidden units and four FC layers with 80 hidden units. 60M timesteps of artificial agent data and 0.2M data acquired from one human player were used for the model training. When the same data were exploited for five epochs, our model training was about nine times faster than the LSTM-FC Net (2 hours vs 18 hours).

Nine participants between 22 to 29 years of age (mean age=25.33) were recruited for the user test. All participants were provided with sufficient practice time to avoid their skill increase during the user test. After the practice time, participants took apart in three sessions with the three different types of DDA agents in random order. The initial difficulty adjustment of each session was performed using data acquired during a pre-session of about one minute performed immediately before each session. Each session lasted about four minutes, and after half of each session (i.e., after two minutes), a short break was given and an additional difficulty adjustment implemented.

\section{Results}
We evaluated the DDA methods using both objective and subjective evaluation metrics. As objective metrics of how successfully the DDA model adapted to the user, we measured the participants' win/loss rate and puck possession, i.e., the percentage of time with puck on one's side. An even game is expected to result in 50 percent for each metric. Figure~\ref{fig:2}(a) indicates that our method shows a comparable win/loss rate to the conventional method and is superior to that of the LSTM-FC Net. In terms of the puck possession, our method also shows a comparable result to the conventional method, and a superior result to the LSTM-FC Net.

As subjective metrics, we asked the participants to complete a questionnaire which assessed the \textit{enjoyment}, \textit{suitable difficulty}, \textit{engrossment}, and \textit{personal gratification}, modified from~\cite{kusano2019motion}. Figure~\ref{fig:2}(b) shows the subjective evaluation results of our user test. Our DDA method shows superior results to the LSTM-FC Net in terms of the \textit{enjoyment}, \textit{engrossment}, and, in particular, the \textit{suitable difficulty} score.

\section{Conclusion}
In this paper, we proposed a novel DDA method named fast user adaptation based on a meta-learning algorithm. Our method surpassed a deep neural network-based baseline in both objective and subjective evaluations, and showed a much faster learning speed. In addition, our method showed comparable performance to the conventional DDA even though it has the advantage of not requiring time-consuming parameter tuning.

\section{Acknowledgments}
This work was supported by the Basic Science Research Program through the National Research Foundation of Korea (NRF) funded by the Ministry of Education (NRF-2018R1D1A1B07043580).

%
%
%
%
%

\balance{}

\bibliographystyle{SIGCHI-Reference-Format}
\bibliography{reference}

\end{document}